# Activity networks determine project performance

Alexei Vazquez*[1], Iacopo Pozzana[1], Georgios Kalogridis[1], Christos Ellinas*[1]

[1]Nodes & Links Ltd, Salisbury House, Station Road, Cambridge, England, CB1 2LA

*corresponding author

# Abstract

Projects are characterised by activity networks with a critical path, a sequence of activities from start to end, that must be finished on time to complete the project on time. Watching over the critical path is the project manager's strategy to ensure timely project completion. This intense focus on a single path contrasts the broader complex structure of the activity network, and is due to our poor understanding on how that structure influences this critical path. Here, we use a generative model and detailed data from 77 real world projects (+$10bn total budget) to demonstrate how this network structure forces us to look beyond the critical path. We introduce a duplication-split model of project schedules that yields (i) identical power-law in- and-out degree distributions and (ii) a vanishing fraction of critical path activities with schedule size. These predictions are corroborated in real projects. We demonstrate that the incidence of delayed activities in real projects is consistent with the expectation from percolation theory in complex networks. We conclude that delay propagation in project schedules is a network property and it is not confined to the critical path.

# Introduction

Delivering projects on time and on budget is necessary to improve human prospect[1], with the World Bank stating that 22% of the world's gross domestic product - about $48 trillion - relies exclusively on project-based delivery mechanisms[2]. Yet the majority of public and private large-capital projects are completed late and over budget[3]. An industry survey captures the scale of the problem - reviewing 10,624 projects from 200 companies in 30 countries and across a variety of industries, it concludes that only 2.5% of projects were completed on time and budget[4]. A recent review reaffirms the stubbornness of the challenge, with delays remaining at comparable levels even after 15 years of project management advancements (comparing projects started between 1998-2003 vs 2013-2018)[4].

This consistency in poor performance suggests that the core method of evaluating delay risk is inadequate for the complex nature of modern projects.

Since the 1960s project managers have almost exclusively relied on monitoring the critical path as the means to manage delay risk. This path is essentially a sequence of activities from start to end that are executed without any slack time in between [5,6].

An increase in the duration of any activity in the critical path causes project end overrun. It is a simple concept and it provides a simple solution: the critical path must be executed as planned at all costs. Yet, modern projects are more complex, with schedules that look like complex networks of activity dependencies [7-8]. Delays in activities outside the critical path can similarly cause project end overruns through domino-like cascades, similar to how viruses spread[9].

Given the consistency in project delays over the past decades, we examine the limit of applicability for the critical path using both synthetic and real data. We find that, beyond a certain level of complexity, the critical path becomes irrelevant and project end overruns are primarily driven by activities that are outside of that path.

# Results

## Generative model of project schedules

Project schedules (or activity networks) can be seen as the outcome of a growth process, where a parent activity can be duplicated or split (Fig. 1A). Generic activities can be duplicated and broken into two smaller activities that run in parallel, both inheriting all predecessors and successors of the parent activity. Specialised activities can be split into two activities executed in sequence, such that one specialised contractor executes the first part and another the second.

Starting from two activities executed in sequence (Fig. 1B), we can grow the network by a stochastic sequence of duplication and split events, with a probability of duplication $q$. For small $q$, activities will be mostly split, generating a mostly linear activity network. For large $q$, most activities are duplicated, leading to a network with numerous parallel paths (Fig. 1).

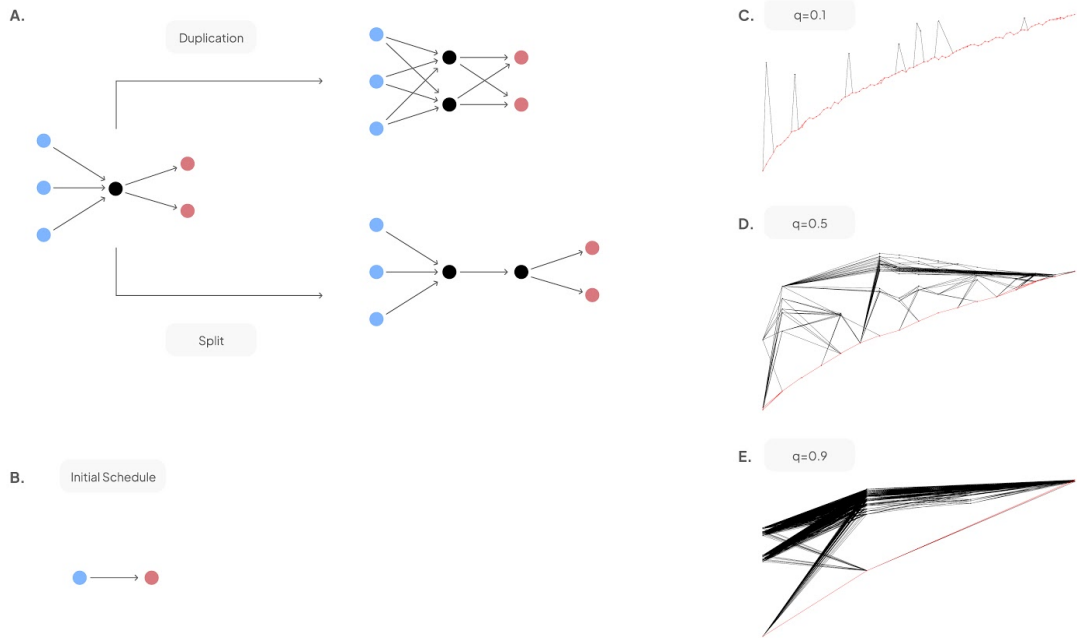

**Figure 1. Duplication-split model.** A) Duplication and split rules acting on an activity (black). B) Initial conditions where the duplication-split model is applied. C) Examples of activity networks generated by the duplication-split model for three different values of the duplication probability $q$. Critical path is highlighted in red.

Node duplication, also known as copying, has been studied in the context of web networks and protein interactions networks[10-13]. It has been shown that node duplication generates networks with a power law probability distribution in the number of links associated to a node[10-13]. In the Methods section we demonstrate that this is indeed the case for our model of duplication-split activity networks, but with a twist. We can show that the distributions of the number $k$ of predecessors and successors to an activity follow the same power law $p_k \sim k^{-1/q}$, where $p_k$ is the probability that an activity has $k$ predecessors (or successors). Our calculations are validated by numerical simulations of the duplication-split model (Fig. 2).

Once we create activity networks, we can populate synthetic project schedules by assigning durations to each activity. We now have project schedules with a critical path, a sequence of activities from the start to the end of the project. The latter carry as a consequence that delaying the finish of any activity in the critical path delays the project end date by the same amount

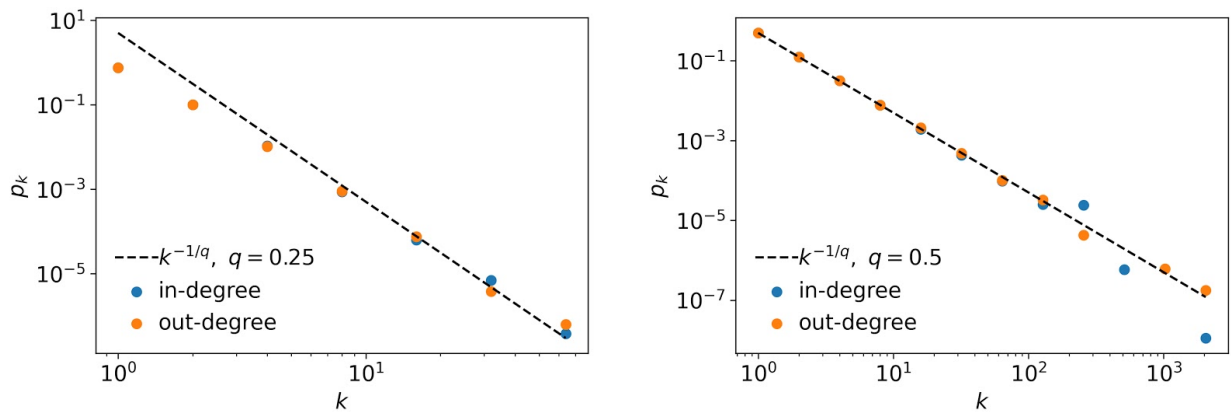

**Figure 2. Degree distributions.** Distribution of the number of predecessors (in-degree) and successors (out-degree) across activities in networks generated by the duplication-split model, using A) $q$=0.25 and B) $q$=0.5. The dashed line highlights the power law tail predicted by our calculations.

## Shrinkage of the critical path

Critical path is the perceived centrepiece in project management due to its sensitivity to delays. Yet, a look at the synthetic activity networks in Fig. 1C made us question whether that critical-path-centric view is valid for modern projects, given that modern projects have complex structures with many parallel paths of work happening at the same time[6].

In cases where activity networks are quasi-linear, the critical path is indeed the dominating structural feature (Fig. 1C, $q$=0.1). In contrast, in the $q$=0.9 activity network we observe a large number of parallel paths with similar number of activities (Fig. 1C, $q$=0.9). It is in these cases that the concept of the critical path may be of less relevance to manage the delay risk of the project.

Following these qualitative observations, we show that the larger the project network, the smaller the relative size of the critical path. Furthermore, the larger the duplication probability $q$, the smaller the relative number of activities in the critical path, in agreement with the visual inspection of the $q$=0.1 and 0.9 synthetic activity networks in Fig. 1C. We determine the number $c$ of critical path activities in a network of $n$ activities and duplication parameter $q$. We estimate that $c \sim n^{1-q}$ and therefore the fraction of activities in the critical path decreases as $c/n \sim n^{-q}$. Numerical simulations corroborate the $c/n \sim n^{-\alpha(q)}$ scaling, albeit with $\alpha(q) \leq q$ (Fig. 3).

We note that the duplication-split networks are not small-world networks[14,15]. In small-world networks the typical distances between nodes scale logarithmically with network size ($c \sim \ln n$)[15]. Duplication-split networks are a new class of networks with power-law degree distributions and power-law scaling of node distances with network size. In fact, these are fractal networks ($c \sim n^{1/D}$), with a fractal dimension $D$=1/(1-$\alpha(q)$).

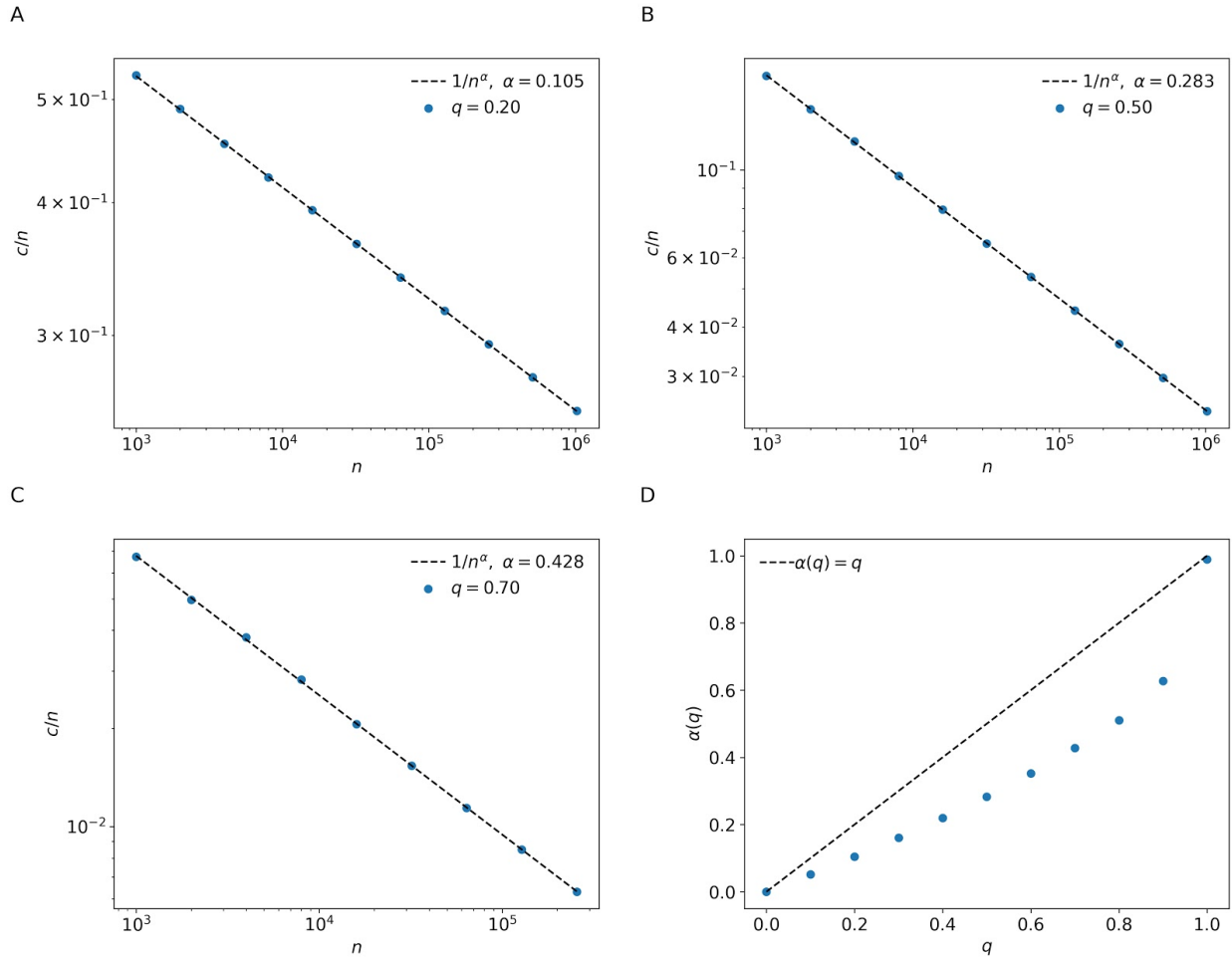

**Figure 3. Relative size of the critical path.** A-C) The number of activities $c$ in the critical path relative to the total number of activities $n$, for activity networks generated by the duplication-split model. The dashed line highlights the power law decay predicted by our calculations. D) The scaling exponent of the power law decay $c \sim n^{-\alpha}$ as a function of $q$ (symbols), The dashed line is the theoretical upper-bound.

## Vanishing critical path in empirical activity networks

To demonstrate that our observations are representative of the real-world challenge, we shift our focus to empirical data for 77 construction projects (total value +$10bn), with activity

networks representing different stages of the project lifecycle, adding up to 323 project schedules. These activity networks vary in size, from 100 to 16,000 activities.

Driven by our synthetic schedule analysis, our prediction that the relative size of the critical path decreases as the number of activities increases is further confirmed in the empirical data.

First, we corroborate the distribution of the number of predecessors (in-degree) and the number of successors (out-degree) to an activity are almost identical and they follow a power law decay (Fig. 4A). Assuming the power-law decay of the duplication split model $p_k \sim k^{-1/q}$, we obtained a maximum likelihood estimate $q$ from the distribution of the number of predecessors and independently from the number of successors. The duplication-split model predicts that the two should coincide. Indeed, the data for the construction projects fall at or in the vicinity of the equality line (Fig. 4B). Furthermore, the duplication $q$ index of real projects is distributed between 0.1 and 0.5, with most values between above 0.2 (Fig. 1C).

Second, we tested the $c/n \sim n^{-\alpha}$ scaling of the fraction of activities in the critical path. The fraction of activities in the critical path $c$ of real activity networks decreases as the number of activities $n$ increases (Fig. 4D, blue symbols). This decrease approximately follows the scaling $c/n \sim n^{-\alpha}$ with $\alpha$=0.79.

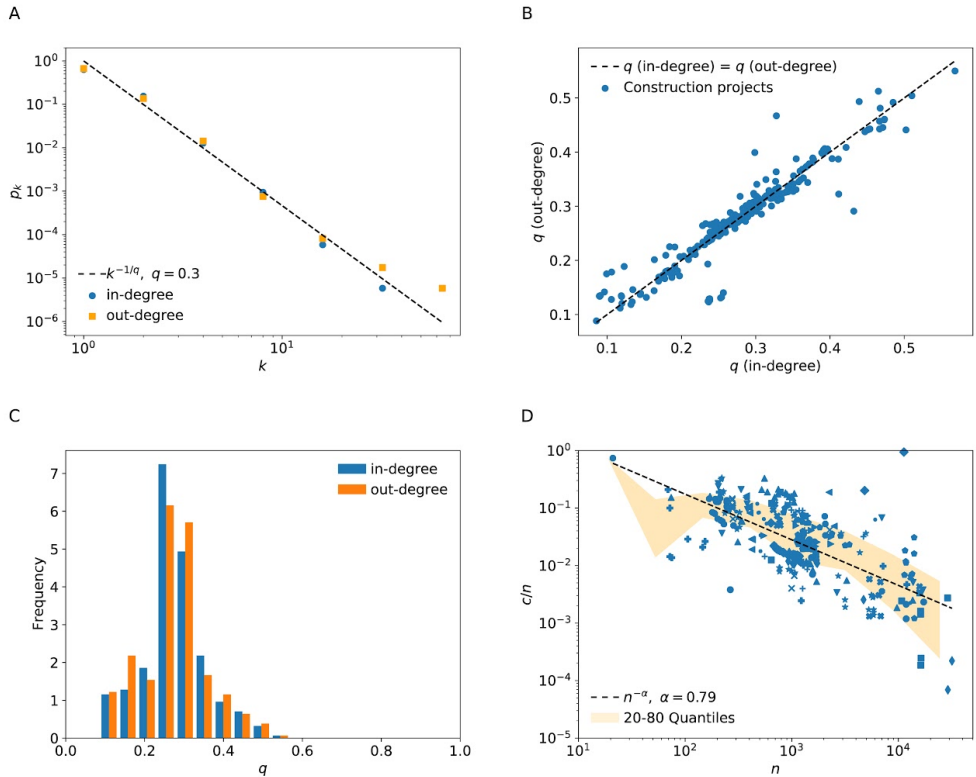

**Figure 4. Empirical data analysed.** A) Distribution of the number of predecessors (in-degree) and successors (out-degree) across activities of a typical construction project. The dashed line highlights a power law fitting the data. B) The exponents $q$ obtained from the fit to the distribution of the number of predecessors $q(in\text{-}degree)$ and successors $q(out\text{-}degree)$. Each point represents a project schedule. The dashed line is the equality line. C) The distribution of estimated duplication rate $q$ across projects. D) The fraction of activities in the critical path as a function of the number of activities. Different symbols represent a project schedule at different stages of completion. The solid background represents the [20,80]% confidence interval.

# Network complexity drives delay risk

Now we switch our attention to delay propagation in activity networks. Exogenous delays such as extreme weather events, pandemics or financial crises can cause some activities to be delayed beyond their planned finish date. When activity delays exceed the spare time between activities (free floats) they propagate downstream triggering a delay cascade. We view activity delays exceeding the free floats as microscopic events and the delay cascades reaching the project end as the macroscopic behaviour. The microscopic events are quantified by the probability $p$ that an activity dependency will transmit a delay. The macroscopic behaviour is quantified by the fraction of activities where the activity delay exceeds its total float. We call the latter the delay incidence.

If the critical path is a key delay risk factor, then the incidence of delay across activities should increase with increasing $p \times c$, where $c$ is the critical path size as denoted above. However, when we plot the delay incidence vs $p \times c$ we actually observe a negative non significant correlation (Pearson correlation coefficient = -0.1, significance = 0.7). Therefore, the delay risk is not determined by the critical path size.

If the critical path vanishes for large projects, and we know that almost all complex projects are delayed, where does this risk come from? After ruling out the standard hypothesis (critical path) we shift our focus to activities outside the critical path.

We use percolation theory as a framework to help us quantify the propensity of the project to exhibit a delay, driven by delays at the activity level[16-18]. Bond percolation indicates that when $p$ exceeds a critical threshold $p_c$ delay cascades will take place with a finite probability. For directed networks with uncorrelated in-degrees and out-degrees $p_c=1/<k>$[18], where $<k>$ is the average out-degree. Percolation theory predicts a phase transition from no macroscopic cascades when $p<p_c$ to a finite risk of macroscopics cascades for $p>p_c$.

This is exactly what we observe for real project networks (Fig. 5B), highlighting that project end overruns are indeed a property of the whole network. For $p<p_c$ the delay incidence is below 1%, almost no risk of project delay. In contrast, for $p>p_c$ the delay incidence increases gradually, and in some cases impacting 15% of the entire project. We note that for some projects with $p>p_c$ the delay incidence is below 1% and the confidence interval reaches zero (Fig. 5B, orange band, $p-p_c>0$). This is expected from percolation theory. The occurrence of macroscopic events is probabilistic. What is different from zero is the probability that such macroscopic events occur.

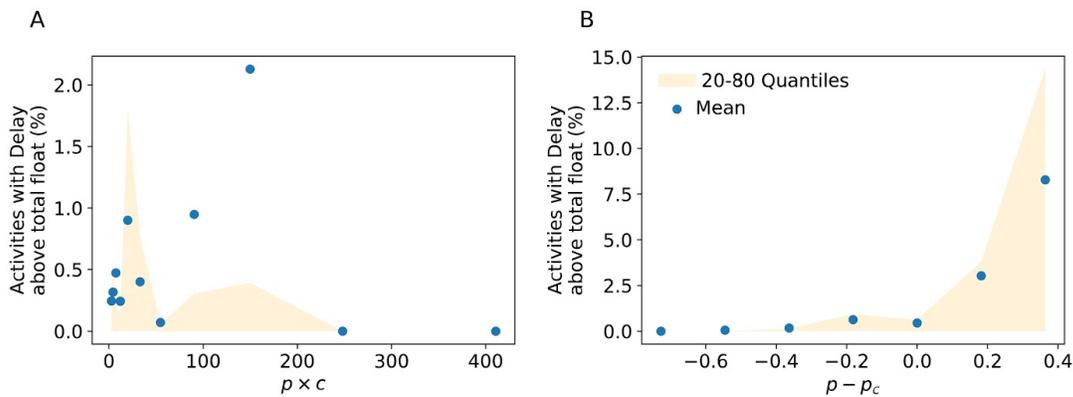

**Figure 5. Delay incidence.** A) Observed delay incidence in construction projects as a function of $p×c$, control parameter associated with the critical path. B) Observed delay incidence in construction projects as a function of $p-p_c$, the control parameter of percolation theory. $p$ is the fraction of delay transmissions along direct activity-activity relations, $c$ the number of activities in the critical path and $p_c$ is the critical threshold from percolation theory ($p_c = 1/<k>$).

## Conclusions

We focus on activity networks that describe large-capital projects, showing that their broader structure contains information about their propensity for delays. Our first contribution is the introduction of the duplicate-split model, and the fact that the duplication index $q$ is a core feature of activity networks. Networks with small $q$ are indicative of quasi-linear topologies, and a good fit for using the critical path. Large $q$ indicates a complex project, where the critical path is relatively small, and parallel paths tend to dominate the overall structure. We then use synthetic and empirical data to both validate the output of the duplicate-split model. Our second contribution is showing that the number of activities in the critical path decreases as $n^{-\alpha}$ and therefore the critical path vanishes in the limit of large activity networks. As a result, the critical path is of limited applicability when it comes to large and complex projects. Our third contribution is the application of percolation theory in order to go beyond the limitations of critical path analysis, whilst showcasing that project end overruns are a network property.

# References

1. Jensen, A., Thuesen, C. & Geraldi, J. The Projectification of Everything: Projects as a Human Condition. *Project Management Journal* **47**, 21–34 (2016).

2. Scranton, P. Projects as a focus for historical analysis: surveying the landscape. *History and Technology* **30**, 354–373 (2014).

3. Flyvbjerg, B., Skamris holm, M. K. & Buhl, S. L. How common and how large are cost overruns in transport infrastructure projects? *Transport Reviews* **23**, 71–88 (2003).

4. Park, J. E. Schedule delays of major projects: what should we do about it? *Transport Reviews* **41**, 814–832 (2021).

5. Kelley, J. E. & Walker, M. R. Critical-path planning and scheduling. in *Papers presented at the December 1-3, 1959, eastern joint IRE-AIEE-ACM computer conference* 160–173 (Association for Computing Machinery, 1959).

6. Santolini, M., Ellinas, C. & Nicolaides, C. Uncovering the fragility of large-scale engineering projects. *EPJ Data Sci.* **10**, 1–13 (2021).

7. Ellinas, C. The Domino Effect: An Empirical Exposition of Systemic Risk Across Project Networks. *Production and Operations Management* **28**, 63–81 (2019).

8. Ellinas, C., Allan, N. & Johansson, A. Toward Project Complexity Evaluation: A Structural Perspective. *IEEE Systems Journal* **12**, 228–239 (2018).1.

9. Vespignani, A. Modelling dynamical processes in complex socio-technical systems. *Nature Phys* **8**, 32–39 (2012).

10. Kleinberg, J. M., Kumar, R., Raghavan, P., Rajagopalan, S. & Tomkins, A. S. The Web as a Graph: Measurements, Models, and Methods. in *Computing and Combinatorics* (eds. Asano, T., Imai, H., Lee, D. T., Nakano, S. & Tokuyama, T.) 1–17 (Springer, 1999).

11. Vázquez, A., Flammini, A., Maritan, A. & Vespignani, A. Modeling of Protein Interaction Networks. *CPU* **1**, 38–44 (2003).

12. Pastor-Satorras, R., Smith, E. & Solé, R. V. Evolving protein interaction networks through gene duplication. *Journal of Theoretical Biology* **222**, 199–210 (2003).

13. Chung, F., Lu, L., Dewey, T. G. & Galas, D. J. Duplication Models for Biological Networks. *Journal of Computational Biology* **10**, 677–687 (2003).

14. Watts, D. J. & Strogatz, S. H. Collective dynamics of 'small-world' networks. *Nature* **393**, 440–442 (1998).

15. Amaral, L. A. N., Scala, A., Barthélémy, M. & Stanley, H. E. Classes of small-world networks. *Proceedings of the National Academy of Sciences* **97**, 11149–11152 (2000).

16. Albert, R., Jeong, H. & Barabási, A.-L. Error and attack tolerance of complex networks. *Nature* **406**, 378–382 (2000).

17. Buldyrev, S. V., Parshani, R., Paul, G., Stanley, H. E. & Havlin, S. Catastrophic cascade of failures in interdependent networks. *Nature* **464**, 1025–1028 (2010).

18. Schwartz, N., Cohen, R., ben-Avraham, D., Barabási, A.-L. & Havlin, S. Percolation in directed scale-free networks. *Phys. Rev. E* **66**, 015104(R) (2002).

# Methods

## Estimation of the degree distribution

Let $n_k(n)$ the number of activities with $k$ predecessors in the activity network. As new activities are added $n_k(n)$ changes according to the equation

(1)  $\quad n_k(n+1) = n_k(n) + q\left[\frac{k-1}{n}n_{k-1}(n) - \frac{k}{n}n_k(n) + \frac{1}{n}n_k(n)\right] + (1-q)\delta_{k1}$ .

The first term inside [...] corresponds to activities with $k$-1 predecessors and the duplication of one predecessor with probability $(k-1)/n$, moving to the $k$ predecessors group. The second term inside the [...] is the same but for activities with $k$ predecessors, moving from the $k$ predecessors group. The third term inside [...] is the chance that an activity with $k$ predecessors is duplicated, thus generating a new activity with $k$ predecessors. Finally, the last term in (1) is the creation of a new activity with one predecessor following a splitting event, where $\delta_{k1} = 1$ if $k$=1 and 0 otherwise (Kronecker delta symbol).

Assuming a steady state solution $n_k(n) = np_k$ we obtain

(2)  $\quad p_k = q\left[(k-1)p_{k-1} - kp_k + p_k\right] + (1-q)\delta_{k1}$ .

We can iterate this equation to obtain an expression for all $k$>1 as a function of $p_1$

(3)  $\quad p_k = \frac{q(k-1)}{1+q(k-1)}p_{k-1} = \prod_{s=1}^{k-1}\frac{s}{\frac{1}{q}+s}p_1 = \Gamma\left(\frac{1}{q}+1\right)\frac{\Gamma(k)}{\Gamma\left(\frac{1}{q}+k\right)}p_1$ ,

where Γ(x) is the gamma function. For k>>1 the later equation has the asymptotic behaviour

(4)    $p_k \sim k^{-1/q}$ .

The same reasoning can be repeated using *k* as the number of activity successors. That is, the distributions of the number of predecessors (in-degree) and successors (out-degree) are identical in the $n \to \infty$ limit.

## Estimation of the critical path size

Consider a network schedule with *n* activities and *c* activities in the critical path. As new activities are added, the size of the critical path can increase if a task in the critical path is subject to the split rule. Since the split rule is executed with probability 1-*q* at each activity addition and the probability that the activity selected for splitting is in the critical path is equal to *c*/*n*, then

(6)    $\frac{dc}{dn} = (1 - q)\frac{c}{n}$ .

Integrating this equation we obtain

(7)    $\frac{c}{n} \sim n^{-q}$ ,

This result is an approximation. As the network grows there could be changes in what activities are in the critical path, making the critical path shorter. We conjecture the scaling *c/n∼n^-α(q)*, where α(*q*)≤*q*.

## Python code for the duplication-split model

```
import numpy as np
Import networkx as nx

def duplication_split_digraph(n,p):
    np.random.seed(0)
    G = nx.DiGraph()
```

```
G.add_nodes_from(range(n))
G.add_edge(0,1)
duplicate = np.random.random(n) < p
for i in range(2,n):
      j = np.random.randint(i)
      G.add_edges_from([(i,k) for k in G.successors(j)])
      if duplicate[i]:
            G.add_edges_from([(k,i) for k in G.predecessors(j)])
      else:
            G.remove_edges_from([(j,k) for k in G.successors(j)])
            G.add_edge(j,i)
```

## Generative model simulations

Project schedules are generated in three steps.  1) We generate an activity network by successive application of the duplication/split rules up to we reach *n* activities. At each activity addition we select an activity with equal probability among all current activities in the network, execute the duplication rule with a probability *q* otherwise the split rule. 2) We assign a duration *x* to each activity from a distribution with probability density function *f(x)*. Here we use an exponential distribution with mean 1 day. 3) We assume that all activity relations are of the standard Finish-Start type, that all activities with no predecessors start at day 0 and apply forward/backward passing [6-7] to determine the early and late start and end dates for all other activities. Average statistics and distributions are estimated from 100 simulations of these steps for each set of parameters (*n,q*).

## Critical path

Once a schedule has been generated, we perform a second backward pass to calculate the total float of each activity. The total float is defined as the amount of time that the end date of an activity can be postponed without affecting the project end date [6-7]. The critical path is the set of activities with total float 0 and it will be denoted by *C*. The size of *C*, the number of activities in the critical path, is denoted by *c*.

## Probability of delay transmission

We estimate the probability *p* that an activity dependency will transmit delays by looking at all completed activities, and computing the fraction of dependencies with slack time that is smaller than the delay at the parent activity, across all relations out-going from finished activities.

## Control parameter of the critical path method

The probability that there are no delay transmissions in the critical path is $P(p,c)=1-(1-p)^c$. For small *p* it can be approximated by $P(p,c)\approx1-e^{-pc}$. This later equation shows that the delay risk associated with the critical path should decrease with increasing *pc*.

## Empirical data of construction projects

The dataset is composed of 77 construction projects, with multiple project schedules for each construction project, totalling 323 project schedules. The project schedules were generated by the project managers using an industry standard enterprise software package (Oracle Primavera P6).

# Endnotes

## Acknowledgements

This work was partly supported by the European Union's Horizon 2020 and the Cyprus Research Innovation Foundation under the SEED program (grant agreement 0719(B)/0124). Nodes & Links Ltd provided support in the form of salary for Alexei Vazquez, Iacopo Pozzana, Georgios Kalogridis and Christos Ellinas, but did not have any additional role in the conceptualisation of the study, analysis, decision to publish, or preparation of the manuscript.

## Authors contributions

Alexei Vazquez conceived the duplication-split model, estimated the model properties, performed the generative model simulations and analysed the data. Iacopo Pozzana and Georgios Kalogridis created and curated the construction project dataset. Christos Ellinas directed the work and provided expertise in project scheduling. Alexei Vazquez and Christos Ellinas wrote the paper.

## Data Availability and Code Availability

All the data necessary to support our conclusions is reported in the figures. Code for the duplication-split model is provided in the methods. Raw data for construction projects has restricted access and can be provided upon consultation.

## Competing interests

The authors declare no competing interests. Nodes & Links provided support in the form of salaries, but did not have any additional role in the study design, analysis, decision to publish, or preparation of the manuscript.